\documentstyle[11pt,epsf]{article}

\makeatletter \@addtoreset{equation}{section} \makeatother

\def\ben{\begin{equation}}
\def\een{\end{equation}}

  \let\n=\nu

\let\C=\Chi

 \def\bd{\begin{document}} \def\ed{\end{document}}
\def\ds{\documentstyle} \let\fr=\frac \let\bl=\bigl \let\br=\bigr
\let\Br=\Bigr \let\Bl=\Bigl
\let\bm=\bibitem
\let\na=\nabla
\let\pa=\partial \let\ov=\overline
\newcommand{\be}{\begin{equation}}
\newcommand{\ee}{\end{equation}}
\def\ba{\begin{array}}
\def\ea{\end{array}}
\def\ft#1#2{{\textstyle{{\scriptstyle #1}\over {\scriptstyle #2}}}}
\def\fft#1#2{{#1 \over #2}}
\def\del{\partial}
\def\vp{\varphi}
\def\sst#1{{\scriptscriptstyle #1}}
\def\oneone{\rlap 1\mkern4mu{\rm l}}
\def\td{\tilde}
\def\wtd{\widetilde}
\def\ie{\rm i.e.\ }
\def\dalemb#1#2{{\vbox{\hrule height .#2pt
        \hbox{\vrule width.#2pt height#1pt \kern#1pt
                \vrule width.#2pt}
        \hrule height.#2pt}}}
\def\square{\mathord{\dalemb{6.8}{7}\hbox{\hskip1pt}}}
\newcommand{\ho}[1]{$\, ^{#1}$}
\newcommand{\hoch}[1]{$\, ^{#1}$}
\newcommand{\bea}{\begin{eqnarray}}
\newcommand{\eea}{\end{eqnarray}}
\newcommand{\ra}{\rightarrow}
\newcommand{\lra}{\longrightarrow}
\newcommand{\Lra}{\Leftrightarrow}
\newcommand{\ap}{\alpha^\prime}
\newcommand{\bp}{\tilde \beta^\prime}
\newcommand{\tr}{{\rm tr} }
\newcommand{\Tr}{{\rm Tr} }
\def\0{{\sst{(0)}}}
\def\1{{\sst{(1)}}}
\def\2{{\sst{(2)}}}
\def\3{{\sst{(3)}}}
\def\4{{\sst{(4)}}}
\def\5{{\sst{(5)}}}
\def\6{{\sst{(6)}}}
\def\7{{\sst{(7)}}}
\def\8{{\sst{(8)}}}
\def\n{{\sst{(n)}}}
\def\cA{{{\cal A}}}
\def\cF{{{\cal F}}}
\def\tV{\widetilde V}
\def\tW{\widetilde W}
\def\tH{\widetilde H}
\def\tE{\widetilde E}
\def\tF{\widetilde F}
\def\tA{\widetilde A}
\def\im{{{\rm i}}}
\def\tY{{{\wtd Y}}}
\def\ep{{\epsilon}}
\def\vep{{\varepsilon}}
\def\R{\rlap{\rm I}\mkern3mu{\rm R}}
\def\bD{{{\bar D}}}

\def\R{\rlap{\rm I}\mkern3mu{\rm R}}
\def\bD{{{\bar D}}}
\def\R{{{\Bbb R}}}
\def\C{{{\Bbb C}}}
\def\H{{{\Bbb H}}}
\def\CP{{{\Bbb C}{\Bbb P}}}
\def\RP{{{\Bbb R}{\Bbb P}}}
\def\Z{{{\Bbb Z}}}
\def\bA{{{\Bbb A}}}
\def\bB{{{\Bbb B}}}
\def\bC{{{\Bbb C}}}
\def\bR{{{\Bbb R}}}
\def\bD{{{\Bbb D}}}
\def\bE{{{\Bbb E}}}
\def\bZ{{{\Bbb Z}}}
\def\cD{{{\cal D}}}
\def\Re{{{\frak{Re}}}}
\def\Im{{{\frak{Im}}}}
\def\cosec{{\,\hbox{cosec}\,}}
\def\Gm{{\Gamma_{\!\! -}}}
\def\Gp{{\Gamma_{\!\! +}}}

\def\stan{{standard }}
\def\nonstan{{supernumerary }}

\def\cosech{{\hbox{cosech}}}

\def\etcyc{{\hbox{and cyclic}}}
\def\btheta{{\bar\theta}}

\newcommand{\tamphys}{\it Center for Theoretical Physics,
Texas A\&M University, College Station, TX 77843, USA}
\newcommand{\umich}{\it Michigan Center for Theoretical Physics,
University of Michigan\\ Ann Arbor, MI 48109, USA}
\newcommand{\upenn}{\it Department of Physics and Astronomy,\\
University of Pennsylvania, Philadelphia,  PA 19104, USA}
\newcommand{\SISSA}{\it  SISSA-ISAS and INFN, Sezione di Trieste\\
Via Beirut 2-4, I-34013, Trieste, Italy}

\newcommand{\mitchell}{\it George P. \& Cynthia W.
Mitchell Institute for Fundamental Physics,\\
Texas A\&M University, College Station, TX 77843-4242, USA}

\newcommand{\newton}{\it Isaac Newton Institute for Mathematical Sciences,\\
0 Clarkson Road,  University of Cambridge, Cambridge CB3 0EH, UK}

\newcommand{\ihp}{\it Institut Henri Poincar\'e\\
  11 rue Pierre et Marie Curie, F 75231 Paris Cedex 05}

\newcommand{\damtp}{\it DAMTP, Centre for Mathematical Sciences,
 Cambridge University,\\  Wilberforce Road, Cambridge CB3 OWA, UK}

\newcommand{\itp}{\it Institute for Theoretical Physics, University of
California\\ Santa Barbara, CA 93106, USA}

\newcommand{\istanbul}{\it Department of Physics, I{\c s}$\imath$k University,  {\c S}ile, Istanbul
34980, Turkey}

\newcommand{\auth}{ R. G\"uven}

\begin{document}
\begin{flushright}

\end{flushright}
\vspace{12pt}\vspace{12pt}\vspace{12pt}\vspace{14pt}
\begin{center}

{\Large {\bf  The Conformal Penrose Limit: Back to Square One }}
\vspace{14pt} \vspace{12pt}

\auth

\vspace{7pt}

\vspace{7pt} \istanbul

\vspace{7pt}

\vspace{14pt} \vspace{14pt}\vspace{14pt}\vspace{12pt} \vspace{14pt}

\underline{ABSTRACT}
\end{center}

    We show that the conformal Penrose limit is an ordinary plane
    wave limit in a  higher dimensional framework which resolves
    the spacetime singularity. The higher dimensional framework is
    provided by  Ricci-flat manifolds which are of the form $M_D = M_d \times B
    $, where  $M_d$ is an Einstein spacetime that has a negative cosmological constant
    and admits a spacelike conformal Killing vector,  and $B$ is a complete Sasaki-Einstein
    space with constant sectional curvature. We define the Kaluza-Klein metric of $M_D$ through the conformal
    Killing potential of $M_d$ and prove that  $M_d$ has a conformal Penrose limit if and only
    if $M_D$ has an ordinary plane wave limit. Further properties of the limit are discussed.

\pagebreak \setcounter{page}{1}

\newpage

\section{Introduction}

   Recently, Penrose limit \cite{pen} and its gauge theory
counterpart, the BMN limit \cite{mald}, have played a pivotal role
in understanding certain aspects of the AdS/CFT correspondence in
string theory. This correspondence has its roots in  $AdS_{p+2}
\times S^{D-p-2} $ type of geometries, which are  products of an
anti de Sitter ($AdS$) spacetime and a sphere $S$ of appropriate
dimensions, and Penrose limits of $AdS_{p+2} \times S^{D-p-2} $
spacetimes were found to be the maximally supersymmetric plane
waves \cite{bla}.  Moreover, in the central $D=10$, $p=3$ case,
superstrings with non-trivial Ramond-Ramond  fields could be
consistently quantized  on the plane wave background \cite{mat}.
Further work based on these developments have furnished us with
new insights about the AdS/CFT correspondence in a framework that
surpasses the supergravity approximation.

Conformal Penrose limit \cite{guv2},\cite{guv3} is a new type of a
limit, taken again in the vicinity of a null geodesic, which allows
two properties to be preserved that were not permitted in the
original Penrose limit. Conformal Penrose limit is designed to
preserve a non-zero cosmological constant $\Lambda$ and also takes
into account the presence of metric functions homogeneous of degree
zero in the coordinates. It turns out that this limit is available
only when the spacetime admits a spacelike conformal Killing vector
and $\Lambda < 0$. Whereas the Penrose limit always yields an
ordinary plane wave, the conformal Penrose limits  of such
spacetimes turn out to be $AdS$ plane waves. These $AdS$ plane waves
can be interpreted as the Randall-Sundrum zero mode \cite{chg},
preserve $1/4$ supersymmetries and possess a Virasoro symmetry
\cite{ban}. The procedure works in all spacetime dimensions $d \ge
4$ but in the case $d=4$, it actually yields no wave degrees of
freedom. This is not suprising because in $d=4$ the only spacetime
that has $\Lambda < 0$ and admits a spacelike conformal Killing
vector (CKV) is the $AdS$ space \cite{gar} and consequently, in this
case the conformal Penrose limit amounts only to a symmetry of a
unique space. This is perhaps the reason why the conformal limit was
not taken into account in the original Penrose argument.

In string theory,  conformal Penrose limit is relevant to the
Freund-Rubin type of compactifications encountered in the study of
AdS/CFT and DW/QFT dualities  in various dimensions \cite{tow},
\cite{berg}. When one considers the corresponding  supergravity
Lagrangians, one finds that the dilaton field must act as a
potential for a CKV in the compactification process. This role of
the dilaton was first utilized in the context  of the  $D=10$
dilatonic branes \cite{duff},\cite{gib} where the field equations
were reduced to the Einstein equations with the appropriate
cosmological constants. In general this type of reduction requires
$\Lambda < 0$ for the resulting lower, $d$-dimensional spacetime,
but does not specify the type of CKV. Remarkably, the Freund-Rubin
compactification does also require the CKV to be spacelike in the
case of the $D=10$, $p=6$ Lagrangian, which is relevant to the
$D6$-branes \cite{guv3}.

Although the $AdS$ plane waves have various desirable properties,
they are also known to suffer from  pp-curvature singularities
\cite{po},\cite{bre}. At this singularity all scalar invariants of
the $AdS$ plane waves are well-behaved, but certain components of
the Riemann tensor, relative to a frame which is parallelly
transported along a causal geodesic, diverges. It is therefore of
considerable interest to see whether the pp-curvature singularity
can be resolved by some means in string theory. This issue was
addressed in \cite{guv3} and it was found that the singularity can
be resolved only in the $D=10$, $p=6$  case by lifting up the
limiting solution to $D=11$ supergravity. Recall that this was the
only case where a condition on the type of the CKV was
encountered.

Remarkably, what one gets in $D=11$ as the oxidation of the
$D=10$, $p=6$ limiting solution is an ordinary plane wave, or  an
asymptotically locally Euclidean (ALE) plane wave with an
$A_{N-1}$ singularity \cite{guv3}. This result raises in turn the
question whether the confomal Penrose limit can be viewed always
as an ordinary Penrose limit in a higher dimensional framework
where the singularity is resolved. The purpose of the present
paper is to furnish the framework in which this expectation is
indeed fulfilled.

For this purpose we shall consider manifolds that are of the form
$M_D = M_d \times B$, where  $M_d$ is a $\Lambda \neq 0$ Einstein
spacetime  that admits a CKV  and $B$ is an internal space of
appropriate dimension.  The Kaluza-Klein (KK) metric of $M_D$ will
be taken to be conformal to the direct product of the metrics of
$M_d$ and $B$. We shall also require $M_D$ to be Ricci-flat in a
conformal gauge where the conformal factor of the metric is
determined solely by the conformal Killing potential of $M_d$. The
treatment will  allow initially both of the signs for the
pseudo-norm of the CKV as well as for $\Lambda$, and we shall see
how spacelike CKV and $\Lambda < 0$ conditions are simultaneously
singled out together with the sign of the Ricci curvature of $B$.
We shall note that almost all $M_d$ of interest are singular at
the fixed point of the CKV and find that these singularities are
always resolved in the corresponding higher-dimensional $M_D$
whenever $B$ is a regular Sasaki-Einstein space. In order to avoid
the presence of a scalar polynomial curvature singularity on
$M_D$, which is not positioned at the fixed point of the CKV, the
internal space $B$ will be further restricted to the complete
Sasaki-Einstein spaces of constant sectional curvature. We shall
prove with this input that each $M_d$ has a conformal Penrose
limit if and only if the corresponding $M_D$ has an ordinary plane
wave limit.

Section 2 contains the proof for the case of hypersurface orthogonal
CKV's. In this case the higher dimensional spacetime turns out to be
of a remarkably simple form: $M_D = N \times C(B)$, where $N$ is the
$(d-1)$-dimensional conformal boundary of $M_d$ and $C(B)$ is the
flat cone over $B$. Due to this structure, a null geodesic of $N$
that is passing from a fixed point of $C(B)$ is a null geodesic of
$M_D$. Taking the Penrose limit of $M_D$ around such a geodesic with
the help of the Penrose coordinates of $N$ and the K\"{a}hler
potential of $C(B)$ gives a plane wave spacetime with at most a
conical singularity. On $M_d$  the same limit is then seen to be a
conformal Penrose limit, giving an AdS plane wave. Section 3
presents the generalization of the argument to the CKV's which
possess a non-zero twist and enables us to conclude that conformal
Penrose limit can always be viewed as an ordinary Penrose limit in a
higher dimension.

\section{The Hypersurface Orthogonal Case}

Let us keep the dimension $D$ arbitrary  and consider spacetimes
that are of the form $ M_D = M_d \times B $, where $B$ is a
$(D-d)$-dimensional Riemannian manifold. We shall use capital
Latin letters $M,N,...$ to label the tensor indices on $M_D$, the
Greek letters $\mu, \nu,...$ will refer to the coordinate bases of
$M_d$ and $m,n,...$ will denote the coordinate indices on $B$. We
shall assume that $d\geq 4$ and the Lorentzian factor $M_d$ is an
Einstein space\footnote{Our spacetime conventions are same as
\cite{guv3}. We use in particular the mostly minus signature on
$M_D$.}:
\be R_{\mu\nu} = [\epsilon(d-1)/\it{l}^2]g_{\mu\nu}, \label{ein} \ee
so that its cosmological constant is
\be \Lambda = [\epsilon (d-1)(d-2)/2\it{l}^2].\ee
Here $\it{l}$ is a real parameter and $\epsilon = \pm 1$ in order
to allow $ \Lambda $ to take both signs.  We shall also demand
that $M_d$ admits a smooth vector field $ V^{\mu} $ satisfying
\be {\mathcal{L}}_V g_{\mu\nu}=2\psi g_{\mu\nu}, \label{ckeq}\ee
where $\mathcal{L}$ is the Lie derivative and $ \psi$ is a
differentiable function on $M_d$. Then it can be deduced from
(\ref{ein}) and (\ref{ckeq}) that $\nabla_{\mu} \psi$  itself must
be a hypersurface orthogonal  CKV on $M_{d}$:
\be \nabla_{\mu} \nabla_{\nu} \psi =  \frac{\epsilon}{\it{l}^2}
\psi g_{\mu\nu}. \label{ck2}\ee

The properties of manifolds  which admit an arbitrary CKV are
well-known \cite{knel}, and around any point with
$\nabla_{\mu}\psi \nabla^{\mu}\psi \neq 0$, one can find a
neighborhood  where the metric $g_{\mu\nu}$ of $M_{d}$ has a
warped product form. In this neighborhood a coordinate system $ \{
y, x^{a}\}$, $ a= 1, ...,(d-1)$, exists where $\nabla_{\mu} \psi =
U(y) \delta^{y}_{\mu}$,  $U= d\psi /dy$ and the line element takes
the form:
\be d{s_{d}}^2 = \eta dy^2 + U^{2}(y) g_{ab}(x) dx^{a}dx^{b}.
\label{wm} \ee
Here $g_{ab}(x)$ is a metric on a $(d-1)$-dimensional manifold $N$
so that $M{_d} = I \times_{U^2} N$, where $I$ is a real interval.
Moreover, $\eta = \pm 1$ is the sign of the pseudo-norm of the
CKV:
\be \nabla_{\mu}\psi \nabla^{\mu}\psi = \eta U^2, \label{tn}\ee
which is in general independent of  the sign $\epsilon$ of the
cosmological constant. Notice that $\eta = \pm 1$ also specifies
whether $N$ is a  Riemannian or a Lorentzian manifold. When
$\epsilon = \eta = -1$, the manifold  $N$ can be viewed as the
conformal boundary of $M_d$ \cite{guv3} .

The metrics which are of the form (\ref{wm}) and satisfy
(\ref{ein}) constitute a two-parameter family of solutions which
is described in detail in \cite{guv3} and it is easy to see that
$M_d$ must be geodesically incomplete for  almost all of these
solutions. For example, the scalar invariant:
\be R_{\mu\nu\kappa\lambda} R^{\mu\nu\kappa\lambda} =
4(d-1)\it{l}^{-4} + U^{-4}\{ 2(d-1)(d-2){(U')^{4}} + 4\eta
(U')^{2} R_{N}  + [R_{abcd} R^{abcd}]_{N} \}, \label{sps} \ee
generically diverges at a zero of $U$ which corresponds to a fixed
point of the CKV. Here $U' = dU/dy$,  the scalar curvature of $N$ is
denoted by $R_{N}$  and in general, a subscript $N$ on a quantity
signifies that it is defined on $N$. The invariant (\ref{sps}) can
be shown to be well-behaved at $U = 0$ if  $ \epsilon = \eta = -1$
and $N$ is taken to be a Ricci-flat manifold whose all scalar
invariants vanish. (A discussion of the structure of such $N$'s can
be found in \cite{col}). However, even in this case $M_d$ will be
incomplete unless $N$ is flat. Since $M_d$ has a conformal Penrose
limit only when $ \epsilon = \eta = -1$, it will be useful to study
this subset in more detail.

 Let us therefore specialize to $ \epsilon = \eta = -1$ and assume that $N$ is a complete,
 Ricci-flat manifold.  Suppose $t^a = dx^a/d\tau$ is the unit tangent to a
 timelike geodesic of $N$ with the affine parameter $\tau$. Let $e^{a}_{j}$ be spacelike unit
 vectors, $j = 1,...,d-2$, such that $ (t^a, e^{a}_{j})$ is an orthonormal basis of $N$,  with the property
 that all $e^{a}_{j}$ are parallelly transported along $t^a$ .
 In order to construct a similar basis for $M_d$, let us next
 introduce  the unit tangent $t^{\mu} = dx^{\mu}/ds$ to a timelike
 geodesic of $M_d$. Then the two affine parameters will be related by $ds/d\tau =
 c_{0}
 U^2$, where $c_{0} $ is a non-zero real constant. Without any loss of generality
 one may choose $c_{0} = 1$ and for this choice it can be checked
 that the set of $d$ unit vectors $(t^{\mu}, e^{\mu}_{j},
 e^{\mu}_{d-1})$ defined by using the data on $N$ as
\bea  t^{\mu} = \{ \frac{1}{U^2} t^a,  \dot{y} \}, & \nonumber
\\
      e^{\mu}_{j} =  \{\frac{1}{U} e^{a}_{j}, 0 \}, & \nonumber
\\
 e^{\mu}_{d-1} =  \{\frac{\dot{y}}{U} t^{a}, \frac{1}{U} \},
      \label{bas}\eea
where $\dot{y} = dy/ds$, is the corresponding orthonormal basis
for $M_d$ which is parallelly transported along $ t^{\mu}$. Here
$y$ is subject to $\dot{y}^2 =  U^{-2} - 1$, because of the
geodesic equation.

If one now examines the components of the Riemann tensor of $M_d$
relative to the basis (\ref{bas}), one finds, for example, that
\be t^{\mu} e^{\nu}_{j} e^{\kappa}_{k} e^{\lambda}_{l}
R_{\mu\nu\kappa\lambda} = \frac{1}{U^3} [t^a e^{b}_{j} e^{c}_{k}
e^{d}_{l} R_{abcd}]_{N},\label{sin} \ee
and since $[t^a e^{b}_{j} e^{c}_{k} e^{d}_{l} R_{abcd}]_{N}$ is
perfectly well-behaved on $N$, it follows that (\ref{sin})
diverges at $ U = 0$ unless $N$ is flat and  this would imply that
$M_d$ = $AdS_d$. Hence for the present class of solutions $M_d$
always suffers at least from a pp-curvature singularity which is
located at a fixed point of the CKV unless $M_d$ = $AdS_d$. As
long as $[R_{abcd} R^{abcd}]_{N} \neq 0$, the singularity is in
fact stronger because, (\ref{sps}) then exhibits a scalar
polynomial singularity at $ U = 0$.

Regardless of the nature of this singularity, each such $M_d$ will
have a conformal Penrose limit whose metric is \cite{guv2}
\be
 d{\hat{s}_{d}}^2 = \frac{\it{l}^2}{z^2}[2 du dv - h_{ij}(u) x^i x^j du^2 - \delta_{ij}dx^i dx^j
- dz^2],\label{met2} \ee
where $z$ is a new coordinate $(0 < z < \infty)$ used in place of
$y$,  the range of the indices $i,j$ is now $i,j =1,2,..., d-3$, and
the metric functions satisfy: $h_{jj} (u) = 0$. (Here and in the
sequel we use hats to distinguish the quantities that are the
endpoints of the limits.) This shows that each such $M_d$ has an AdS
plane wave as a limit and although in general the presence of a
singularity is not a hereditary property in the sense of
\cite{geroch}, in our context it is preserved under the limit. Since
(\ref{met2}) always has a pp-curvature singularity
\cite{po},\cite{bre} at the $z=\infty$ fixed point of the CKV, what
may not be inherited by the conformal Penrose limit is the type of
the singularity.

Returning back to the $D$-dimensional picture, let us suppose that
$ M_D $ is equipped with the metric:
\be
 ds_D^2 = (\ell / \psi)^{2}[ds_d^2 + ds_B^2], \label{Dmet}
\ee
where $ ds_d^2 $ and $ ds_B^2 $ are the metrics of $M_{d}$ and $B$
respectively. Treating initially $ \epsilon $ and $\eta$ as
independent sign indicators, we also require $ M_D $ to be
Ricci-flat :
\be R_{MN} = 0. \label{rf} \ee
It then follows from (\ref{rf}), (\ref{ein}) and (\ref{ck2}) that
$B$ must also be an Einstein space:
\be R_{mn} = [-\epsilon(D-d-1)/\it{l}^2]g_{mn}, \label{einK} \ee
but with a cosmological constant that has an opposite sign, and
that
\be g^{\mu \nu} \nabla_{\mu}\psi \nabla_{\nu} \psi =
\frac{\epsilon}{\it{l}^2} \psi^{2}.\label{type} \ee
Another consequence of our assumptions is that on  $M_{D}$ the
components of the Riemann tensor $R_{MNPQ}$  obey
\be R_{{\mu}m{\nu}n} = 0. \label{Dcon} \ee
Conversely, if one starts from (\ref{rf}), (\ref{Dcon}), treats
$\psi$ in (\ref{Dmet})  as an arbitrary smooth scalar field on
$M_d$ and imposes (\ref{einK}), then the conditions (\ref{ein}),
(\ref{ck2}) and (\ref{type}), which completely specify the type of
$M_d$ are obtained.

When $M_{D}$ is constructed in  this manner from the two-parameter
family of $M_d$, the condition (\ref{type}) together with (\ref{tn})
require that
\be \eta = \epsilon,\label{signs}\ee
and consequently, $\eta$ and $\epsilon$ can no longer be
independent. The same condition also requires
\be \psi^2 = \it{l}^2 U^2, \ee
which is a constraint on the two parameters of the $d$-dimensional
solutions, reducing the available $M_d$ to the subset for which
$N$ is Ricci-flat. It follows that the metrics (\ref{wm}) that can
be uplifted to $M_{D}$ by the above procedure must be of the form
\be d{s_{d}}^2 = \frac{\it{l}^2}{z^2}[ g_{ab}(x) dx^{a}dx^{b} +
\epsilon dz^2 ], \label{fm}\ee
with  $g_{ab}$ satisfying  $ [R_{ab}]_N = 0$. In terms of these
coordinates, $U = \it{l} / z$ and $\psi = \pm \it{l}^2 / z$. When
$\epsilon = -1$ and  $N$ is taken to be the $(d-1)$-dimensional
Minkowski space with $x^a$ denoting the usual Minkowski
coordinates,  (\ref{fm}) reduces to the Poincar\'{e} patch of $M_d
= AdS_d$.

Forming the $D$-dimensional metric (\ref{Dmet}) with this input
then gives
\be
 ds_D^2 =  g_{ab}(x)
dx^{a}dx^{b} + \epsilon dz^2 + z^2 d\Omega^2 , \label{Dmet1} \ee
where we have rescaled the  metric on $B$ as $ds_B^2 = \it{l}^2
d\Omega^2$, and relative to the (negative-definite) metric
$d\Omega^2$ the field equation for $B$ is now:
\be R_{mn} = -\epsilon(D-d-1)g_{mn}. \label{einK1} \ee
This shows that the $D$-dimensional Ricci-flat spacetimes that are
constructed from  Einstein manifolds admitting a CKV by the above
procedure are necessarily of the form:
\be
 M_D = N \times  C(B)
 \ee
where $C(B)$ is the Ricci-flat cone over $B$.
When $\epsilon = -1$, the cone  is Riemannian  whereas $\epsilon =
1$ implies that $C(B)$ is Lorentzian, and in both cases $ z\partial
/\partial z$ is an Euler vector field on $C(B)$ generating an
infinitesimal homothety. Such cones are known to play interesting
roles in the context of AdS/CFT correspondence \cite{gryc},
\cite{afhs}.

One may view the above discussion as a  KK reduction of the
$D$-dimensional Ricci-flat theory to Einstein spaces $M_d$ which
is obviously a consistent reduction \cite{hap}. It would be
desirable to maintain this consistency also in reductions to
dimensions higher than $d$ and a prerequisite for this behavior
would be that $B$ admits Killing vectors. Suppose $B$ is compact
and orientable. Since (\ref{einK1}) must hold, it then follows
from Bochner's argument \cite{boch} that, in order $B$ to have
isometries,
 \be \epsilon = -1. \label{fix} \ee
 Due to this reason from now on
we assume $C(B)$ is a Riemannian cone.

Our next assumption about $B$ is that it is a $U(1)$ bundle over a
$(D-d-1)$-dimensional manifold $K$. This allows us to write
\be d\Omega^2 = {d\bar{\Omega}}^2 - (dY + \bar{A})^2 , \label{D-1}
\ee 
where $Y$ is the Killing coordinate, $\bar{A}$ is the KK potential
one-form and a bar over a quantity means that it is defined on
$K$. When (\ref{fix}) and (\ref{D-1}) are substituted into
(\ref{einK1}), the consistency of the $(D-d-1)$-dimensional
equations requires that $K$ is K\"{a}hler  and $\bar{F} =
d\bar{A}$ is related to the K\"{a}hler form $\bar{w}$ of $K$ by
$\bar{F} = 2 \bar{w}$. One then sees that the line element
${d\bar{\Omega}}^2$  must obey
\be \bar{R}_{\alpha \beta} = (D-d+1) \bar{g}_{\alpha \beta}, \ee
where $\alpha, \beta, ..$ are the tensor indices on  $K$. Hence
$K$ must be an even-dimensional, K\"{a}hler-Einstein manifold with
positive Ricci curvature to maintain consistency. This result in
turn implies that $B$ must be a regular Sasaki-Einstein manifold
and consequently, the metric cone $C(B)$ is not only Ricci-flat
but must also be K\"{a}hler, i.e. a Calabi-Yau cone. The
properties of Sasaki-Einstein manifolds and their K\"{a}hler cones
have been extensively studied \cite{boga}, \cite{msy}. It is known
in particular that $ \xi= J(z\partial/\partial z)$, where $J$ is
the complex structure on $C(B)$, is the Reeb vector field which is
both holomorphic and Killing. The one-form  $dY + \bar{A}$ is the
contact form of $B$ and is the dual to the vector field $\xi$.
Moreover, $ z^2$ can be interpreted \cite{msy} as the K\"{a}hler
potential of $C(B)$.

Assuming that $N$ is complete, it is manifest in (\ref{Dmet1})
that the $z = \infty$ singularity of $M_d$ is resolved on $M_D$ .
Unless $B$ is taken to be an odd-dimensional unit sphere with the
round metric, what one now has in the $D$-dimensional picture is a
singularity at $z = 0$ whose nature depends crucially on the
curvature of $B$. In the $d$-dimensional picture, $z = 0$ is the
locus of the conformal boundary $N$ of $M_d$ and is perfectly
well-behaved. The corresponding $M_D$, however, suffers there from
a scalar polynomial curvature singularity if the curvature of $B$
is not constrained. One finds, for example, that the invariant:
\be R^{MOPQ} R_{MOPQ} = [ R^{abcd} R_{abcd} ]_{N} + (\ell / z)^{4}
[ R^{mnpq} R_{mnpq} - 2 \ell^{-4} (D-d)(D-d-1) ], \ee
diverges at $z = 0$ if the curvature of $B$ does not render the
second term to zero. One way to avoid the presence of scalar
polynomial singularities on $M_D$ is to demand that the internal
space $B$ has the minimum non-trivial dimension. In three dimensions
the universal cover of $B$ is isomorphic to the standard
Sasaki-Einstein metric of $S^3$ and $C(B)$ is always a flat cone.
This situation is precisely what was encountered in the framework of
$D=11$ supergravity theory \cite{guv3}. More generally, the same
requirement can be met by specializing to $B$ that are complete and
have constant sectional curvature. Killing-Hopf theorem then implies
that
\be
 B = {S^{D-d}} / \Gamma, \label{int}
\ee
where $\Gamma$ is a  freely acting discrete subgroup of
$O(D-d+1)$. With the choice (\ref{int}) all curvature invariants
of $M_D$ reduce to those of $N$ and since $C(B)$ is again a flat
cone, one has at most a conical singularity at $z = 0$ . It is
known that if $B$ is complete, then $C(B)$ is either flat or has
irreducible holonomy \cite{gall} and the absence of scalar
polynomial curvature singularities leaves out many interesting
Sasaki-Einstein spaces when the dimension is not minimal. When $B$
is the round unit sphere $M_D$ has, of course, no singularity.

Consider now the ordinary Penrose limits of $M_D$. Since $M_D = N
\times  C(B)$, the set of all null geodesics of $M_D$ can be
viewed as the union of two disjoint subsets. In the first subset
one has the null geodesics of $N$ that are passing from fixed
points of $C(B)$ and the second subset is composed of the null
geodesics which have one-dimensional traces on $C(B)$. The second
subset can be viewed as the geodesics of $C(B)$ plus the timelike
geodesics of $N$. Suppose we choose a null geodesic from the first
subset and apply the Penrose limit to its neighborhood. Then the
Penrose coordinates of $N$ together with the K\"{a}hler potential
of $C(B)$ are sufficient to specify the $D$-dimensional scaling
rules. In addition to the standard Penrose scalings \cite{pen} on
$N$, what one needs is to impose that the K\"{a}hler potential of
$C(B)$ scales according to
\be z \rightarrow \Omega_{0} z, \ee
where $\Omega_{0}$ denotes the scaling parameter. Conformally
rescaling  the metric of (\ref{Dmet1}) as $\breve{g}_{MN} =
\Omega_{0}^{-2} g_{MN}$ and  taking the limit $\Omega_{0}
\rightarrow 0$ then gives
\be
 d{\hat{s}_D}^2 = 2 du dv - h_{ij}(u) x^i x^j du^2 - \delta_{ij}dx^i dx^j - dz^2 + z^2 d\Omega^2 , \label{Dmet0} \ee
which shows that the limiting spacetime is  a particular
$D$-dimensional plane wave spacetime for which the wave degrees of
freedom of $M_D$ coincide with that of $N$ and a conical singularity
at $z = 0$ is allowed.

Notice that in (\ref{Dmet1}) the dependence on the parameter $\ell
$ has completely disappeared. This is to be interpreted as a
conformal gauge choice. Since the Ricci-flatness condition is
preserved under the homotheties, it is clear that $\ell $ can
appear as a constant conformal factor in other conformal gauges.
In the $D$-dimensional picture its scaling rule:
\be
\ell \rightarrow \Omega_{0} \ell, \ee
can be inferred  in the chosen conformal gauge  by demanding that
the conformal factor $\ell / \psi$ of (\ref{Dmet}) remains
invariant under the Penrose scalings .

 Since $M{_d} = I \times_{U^2} N$, the null geodesic of $N$ that was
used to reach (\ref{Dmet0}) can be viewed also as a a null geodesic
of $M_d$ which is passing from a fixed point of $I$. Taking the
conformal Penrose limit around such a geodesic of $M_d$ involves
precisely the same scalings that were employed on $M_D$. We
therefore conclude that $M_d$ has a conformal Penrose limit
(\ref{met2}) if and only if $M_D$ has the plane wave limit
(\ref{Dmet0}).

\section{Inclusion of the  Twist of the CKV}

In this section we wish to consider a generalization of the above
discussion which takes into account the presence of a non-zero
twist of the CKV. For this purpose the coordinate system of
(\ref{wm}) is not a suitable starting point. We therefore proceed
as in \cite{guv2} and utilize the fact that  one can locally find
another metric $ \tilde{g}_{\mu\nu}$ on $M_d$ which is conformal
to the original metric of (\ref{ein}):
\be g_{\mu\nu} = W^{-2}\tilde{g}_{\mu\nu}, \label{ct} \ee
and for which $V^\mu$ is an ordinary Killing vector:
${\mathcal{L}}_V {\tilde{g}_{\mu\nu}} = 0 $. Here  $ W $ is a
differentiable scalar field and the  map (\ref{ct}) will be
available as long as $V^\mu$ has no fixed points in the
neighborhood. Choosing the Killing coordinate as $V^{\mu} =
\delta^{\mu}_{z}$ and using the standard  KK decomposition one can
express the line element for $ \tilde{g}_{\mu\nu}$ in the from
\be d \tilde{s} ^2_{d} = g_{ab}(x^c) dx^a dx^b + \eta \lambda^{2}
( dz + \zeta)^{2} ,
\ee 
where $ x^a $ are the remaining coordinates, $\eta
\lambda^{2}(x^c) = \tilde{g}_{\mu\nu} V^{\mu}V^{\nu}$ so that
$\eta$ is again the sign of the pseudo-norm of the CKV and $\zeta
= \zeta_{a} dx^a $ is a KK one-form. The CKV will have a non-zero
twist if and only if $(dz + \zeta) \wedge d\zeta \neq 0$.

Let us assume that  the conformal factor satisfies, with respect
to the new metric, the conditions:
\be \tilde{\nabla}_{\mu} \tilde{\nabla}_{\nu} W = 0, \hspace{.3in}
 \tilde{g}^{\mu \nu} \tilde{\nabla}_{\mu} W  \tilde{\nabla}_{\nu} W = \frac{\epsilon}{\it{l}^2}, \label{cf}\ee
which ensure that $\tilde{g}_{\mu\nu}$ is a Ricci-flat metric on
$M_d$. Equivalently, these conditions imply that $\nabla_{\mu}
W^{-1}$ is a closed CKV for the Einstein metric $g_{\mu\nu}$. It is
therefore possible to identify
\be \psi = \ell W^{-1}, \label{cpot}\ee
and check whether $\psi $ is related to the conformal Killing
potential $ \psi_ {V} = \nabla_{\mu} V^{\mu}/ d$ that is associated
with $V^\mu$.

The equations (\ref{cf}) have the simple solution
\be W = z / \ell + \chi (x^{a}, \ell ),\label{W} \ee
provided $ k_{a} = \zeta_{a} - \ell  \nabla_{a} \chi$ is a Killing
vector for the $(d-1)$-dimensional metric $g_{ab}$ and
\be \eta \lambda^{-2} +  k^{a} k_{a} = \epsilon .
 \label{arr} \ee
Here $k^a k_a = g_{ab} k^a k^b$ and $\chi(x^{a}, \ell )$ is an
arbitrary differentiable function which may possess terms that are
homogeneous of degree zero in $\ell$ and $x^a$.  For this solution
$\psi_ {V}= -\psi/{\ell^2}$ and using $\psi_ {V}$ in (\ref{Dmet})
rather than (\ref{cpot}) only amounts to working in another
$\ell$-dependent conformal gauge.

From (\ref{arr}) it follows that, in order to allow a specialization
to $ k_a = 0 $ (or to $k^a k_a = 0$) in the relevant solutions, one
must require $\eta = \epsilon$. When (\ref{cf}) holds and the metric
of $M_D$ is constructed according to (\ref{Dmet}) and (\ref{cpot}),
the field equation (\ref{rf}) continues to imply (\ref{einK}). By
the same assumptions on $B$ one again ends up with (\ref{fix}) and
the new form of the metric on $M_d$ does not, of course, alter the
conclusion that $B$ is a Sasaki-Einstein space. After taking these
considerations into account and redefining $ z + \ell \chi$ as a new
$z$ coordinate, the Einstein metric of $M_d$ can be cast into the
from
\be
 d{s_d}^2 = \frac{\it{l}^2}{z^2}[ g_{ab}(x)
dx^{a}dx^{b} - \lambda^{2} ( dz + k )^{2} ], \label{dmett} \ee
where $ k = \zeta - \ell d\chi$ and the Ricci-flatness of $
\tilde{g}_{\mu\nu}$, or equivalently (\ref{ein}), requires that
\begin{eqnarray}
\nabla_{a}( \lambda^3 f^{ab}) = 0,\nonumber \\
\lambda^3 f^{ab} f_{ab} + 4 \Delta \lambda = 0,\nonumber \\
R_{ab} = 2^{-1}\lambda^2 {f_{a}} ^{c} f_{bc} - \lambda^{-1}
\nabla_{a} \nabla_{b} \lambda, \label{rN}
\end{eqnarray}
where  $f_{ab} = \nabla_{a} k_{b} - \nabla_{b} k_{a} $  and all
quantities including the D'Alembertian $\Delta = \nabla^a
\nabla_a$ again refer to $g_{ab}$. These equations generalize the
$[R_{a b}]_N = 0$ result of the previous section but it should be
noted that $g_{ab}$ is no longer the metric on the conformal
boundary of $M_d$. The boundary $N$ is still located at $z = 0$
but is now equipped with the  metric $[g_{ab}]_N  = g_{ab} -
\lambda^2 k_a k_b$. It turns out  that, if the twist of the CKV is
non-zero,  $[g_{ab}]_N$ also has a non-vanishing Ricci tensor
prior to the limit. The corresponding Ricci-flat metric of $M_D$
is
\be
 ds_D^2 =  g_{ab}(x) dx^{a}dx^{b} - \lambda^{2} ( dz + k )^{2} + z^2 d\Omega^2 ,
\label{Dmet2} \ee
where the line element $d\Omega^2$ of $B$ is  again governed by
(\ref{einK1}). Since $k^a$ is also a Killing vector for
$[g_{ab}]_N$, it is possible to express all the quantities
appearing in (\ref{Dmet2}) in terms of fields defined solely on
$N$ or  $C(B)$. Letting $\lambda_{N} = [g_{ab}]_{N}  k^a k^b$ and
$k_{N} = [g_{ab}]_{N}  k^a dx^b $, one finds that
\be ds_D^2 = ds_N^2 + \lambda_{N}dz^2 - 2k_{N}dz +  ds^{2}_{C(B)},
\ee and consequently, $z^2$ is still the K\"{a}hler potential of
$C(B)$ but the direct product metric form is no longer available.

 The presence of a non-zero twist does not modify the
singularity structures of these manifolds.  An examination of the
invariant $R^{\mu\nu\lambda\kappa} R_{\mu\nu\lambda\kappa} $ of
$M_d$ shows that as long as $\tilde{R}^{\mu\nu\lambda\kappa}
\tilde{R}_{\mu\nu\lambda\kappa}$ is regular and non-zero, $M_d$
suffers from a scalar polynomial curvature singularity at $z =
\infty$. If all curvature invariants constructed from
$\tilde{g}_{\mu\nu}$ turn out to be zero, this should become a
pp-curvature singularity. Regardless of its nature, the singularity
at $z = \infty$ is always resolved in the higher-dimensional $M_D$.
Provided the curvature invariants  of $\tilde{g}_{\mu\nu}$ are
well-behaved, the higher dimensional $M_D$ can be singular only at
$z = 0$ and in order to avoid scalar polynomial curvature
singularities, one must again choose $B = {S^{D-d}} / \Gamma$. With
this choice the curvature invariants of $M_D$  reduce to the
invariants constructed solely from $\tilde{R}_{\mu\nu\lambda\kappa}$
with respect to the metric $\tilde{g}_{\mu\nu}$.

After taking the conformal Penrose limit of $M_d$ by using the
gauge conditions and the scaling rules of \cite{guv2},  the limit
of (\ref{dmett}) can be brought to the form
\be
 d{\hat{s}_{d}}^2 = \frac{\it{l}^2}{{z}^2}[2 du dv - h_{ij}(u) x^i x^j du^2 - \delta_{ij}dx^i dx^j
-\lambda^{2} (dz + \hat{k})^2],\label{met1} \ee
where $\lambda = \lambda(u)$ and $ \hat{k} = (\dot{b}_{j} x^{j} -
\it{l} \dot{c}) du - b_{j} dx^j $ with $b_{j} = b_{j}(u),  c =
c(u)$ and a dot denotes differentiation with respect to the null
coordinate $u$. The field equation (\ref{ein}) and the conditions
(\ref{cf}) are fully satisfied provided that
\be \lambda^{-2} + b_{j}b_{j} =1,  \ee  \be \ddot{b}_{j} =
h_{jk}\:b_k, \ee \be \ddot{c} = 0, \ee and \be h_{jj}=  -
\ddot{\lambda}/\lambda - 2{\lambda}^2 \dot{b}_j \dot{b}_j . \ee
These equations imply that the limit of  $g_{ab}
$ is a plane wave
metric which is not Ricci-flat whereas $N$ is now equipped with a
pp-wave metric that is Ricci-flat.

Since (\ref{met1}) is obtained by using the Penrose coordinates of
$ \tilde{g}_{\mu\nu}$ and the associated null geodesics of $M_d$
correspond to the null geodesics of $M_D$ that are passing from
fixed points of $B$, what has been accomplished in the
$D$-dimensional picture is an ordinary Penrose limit giving
\be
 d{\hat{s}_{D}}^2 = 2 du dv - h_{ij}(u) x^i x^j du^2 - \delta_{ij}dx^i dx^j
-\lambda^{2} (dz + \hat{k} )^2 + z^2 d\Omega^2. \label{met22} \ee
Noting that this is again the metric of a plane wave spacetime
with at most a conical singularity at $z = 0$, we conclude that
conformal Penrose limit can be viewed as an ordinary plane wave
limit in a higher dimension even when the CKV is not hypersurface
orthogonal.

\section{Discussion}
It is well known that the plane waves owe their universal status
as Penrose limits of general spacetimes to the existence of null
geodesics. When one blows up  a conjugate point-free neighborhood
of such a geodesic of a given spacetime uniformly through the
Penrose procedure, a plane wave spacetime results.  In spacetime
dimensions greater than four further care, however, must be
exercised if the initial spacetime admits a CKV as well as a
non-zero cosmological constant.  It has been realized for some
time that if $ \Lambda < 0$ and the CKV is spacelike, there is a
distinguished class of null geodesics on such spacetimes which
allows a more general, conformal Penrose  limit. The neighborhoods
of these geodesics can be blown in such a way that preserves the
$\Lambda \neq 0$ condition and one then ends up with AdS plane
waves. In this paper we have seen that the conformal Penrose limit
can be viewed as an ordinary Penrose limit in a higher dimension.
Conversely, we have found that certain Penrose limits can be
interpreted as conformal Penrose limits in lower dimensions, and
it can be concluded  that the dimensional reduction and oxidation
processes commute with the limiting procedures in the present
framework.

In this framework two crucial roles were played by the conformal
Killing potential $\psi$ and the conformal boundary $N$ of $M_d$.
Assuming that $M_{D}$ = $M_{d} \times B$, we have considered the
whole conformal class of the metrics on $M_D$ and demanded $g_{MN}$
to be Ricci-flat in the conformal gauge of (\ref{Dmet}). In the
context of string theory this gauge choice corresponds to working in
the the dual frame \cite{tow}. Treating initially  the signs of
$\Lambda$ and the pseudo-norm of the CKV as independent and
arbitrary, the higher dimensional framework elucidated why the $
\Lambda < 0$, spacelike CKV case must be singled out. Assuming that
the CKV is hypersurface orthogonal, it is now clear that the
Ricci-flatness of $M_D$ in the chosen conformal gauge forces the
signs of $\Lambda$ and the pseudo-norm of the CKV to be the same,
$M_D$ to have the form $M_D$ = $N \times C(B)$ and $ \Lambda < 0$
ensures that $C(B)$ is a Riemannian, Ricci-flat cone over a
Sasaki-Einstein $B$. In this case $N$ is Lorentzian and therefore
possesses null geodesics which can be elevated to the null geodesics
of either $M_d$ or $M_D$. It is precisely these geodesics which
allow one to map one limit into the other through oxidation or
reduction. When $ \Lambda > 0$ these null geodesics are no longer
available and consequently, conformal Penrose limit can never give
rise to a dS plane wave rather than an AdS plane wave.

In the higher dimensional picture the conformal Killing potential,
more precisely $\ell^{4}\psi^{-2}$, takes the role of the
K\"{a}hler potential of $C(B)$ and thereby allows one to infer the
higher dimensional scaling rules from those of $M_d$. Taking the
ordinary Penrose limit of $M_D$ gives in general the plane wave
limit $\hat{N}$ of $N$ times a Calabi-Yau cone: $\hat{M}_D$ =
$\hat{N} \times C(B)$ and even in this general setting there is a
remarkable dual singularity structure on $M_d$ and $M_D$. The CKV
fixed point singularity of $M_d$ is always resolved on $M_D$ but
now $M_D$ turns out to be  singular at the apex of the Calabi-Yau
cone. The cone singularity of $M_D$ is in turn resolved in the
lower dimensional picture since it just corresponds to the locus
of the conformal boundary of $M_d$. Although it is an interesting
limit on its own, $\hat{M}_D$ with this general form is obviously
not a plane wave spacetime. One requirement to end up with genuine
plane waves in $D$-dimensions would be the vanishing of all the
scalar polynomial curvature invariants in the limit and this was
ensured by specializing to $ B = {S^{D-d}} / \Gamma $. Prior to
the limit, the same specialization had the virtue of eliminating
all the scalar polynomial singularities of $M_D$.

In general a CKV can carry the degrees of freedom coded in its
twist, in addition to its conformal Killing potential, and we have
seen how the presence of a non-zero twist generalizes the cone
structure of $M_D$ and the metric on the conformal boundary of
$M_d$. We have found that a CKV which has all the available degrees
of freedom also allows the same mapping between the conformal and
the ordinary Penrose limits. We feel that the implications of this
mapping to the AdS/CFT and DW/QFT dualities, especially for the
$D$=11, $d$=8 case, merit further investigation. For this case the
underlying framework would be furnished by $D$=11 supergravity
together with the $SU(2)$ gauged, $d=8$ supergravity and ungauged
$d=7$ supergravity theories \cite{guv3}.

\section*{Acknowledgements}

 The research reported in this paper has been supported in part
by the Turkish Academy of Sciences (T\"{U}BA).


\end{document}